\documentclass[lettersize,journal]{IEEEtran}
\usepackage{amsmath,amsfonts}
\usepackage{algorithmic}
\usepackage{algorithm}
\usepackage{array}
\usepackage[caption=false,font=normalsize,labelfont=up,textfont=up]{subfig}
\usepackage{textcomp}
\usepackage{stfloats}
\usepackage{url}
\usepackage{verbatim}
\usepackage{graphicx}
\usepackage{cite}
\usepackage{multirow}
\usepackage{subfig}
\usepackage{xcolor}
\usepackage{verbatim}
\usepackage{booktabs}
\usepackage{arydshln}
\usepackage{bbding}
\usepackage{tcolorbox}

\begin{document}

\title{AlignFormer: Modality Matching Can Achieve Better Zero-shot Instruction-Following Speech-LLM}

\author{Ruchao Fan, Bo Ren, Yuxuan Hu, Rui Zhao, Shujie Liu, Jinyu Li\\

\thanks{R. Fan, B. Ren, Y. Hu, R. Zhao, S. Liu, and J. Li are with Microsoft Corporation, WA, USA (e-mail: \{ruchaofan, boren, yuxuanhu, ruzhao, shujliu, jinyli\}@microsoft.com).}
}



\maketitle

\begin{abstract}
Integrating speech into LLM (speech-LLM) has gaining increased attention recently. The mainstream solution is to connect a well-trained speech encoder and LLM with a neural adapter. However, the length mismatch between the speech and text sequences are not well handled, leading to imperfect modality matching between the speech and text. In this work, we propose a novel neural adapter, AlignFormer, to reduce the length gap between the two modalities. AlignFormer consists of CTC and dynamic-window QFormer layers, where the CTC alignment provides the dynamic window information for QFormer. The LLM backbone is frozen in training to preserve its text capability, especially the instruction following capability. When training with ASR data only, the proposed AlignFormer unlocks the instruction following capability for speech-LLM and the model can perform zero-shot speech translation (ST) and speech question answering (SQA) tasks. In fact, speech-LLM with AlignFormer can theoretically perform any tasks that the LLM backbone can deal with in the speech version. To evaluate the effectiveness of the instruction-following speech-LLM, we propose to use instruction following rate (IFR) and offer a systematic perspective for the IFR evaluation. In addition, we find that the audio position in training would affect the instruction following capability of speech-LLM and conduct an in-depth study on it. Our findings show that audio-first training achieves higher IFR than instruction-first training. The AlignFormer can achieve a near 100\% IFR with audio-first training and game-changing improvements from zero to non-zero IFR on some evaluation data with instruction-first training. We believe that this study is a big step towards the perfect speech and text modality matching in the LLM embedding space.

\end{abstract}

\section{Introduction}
\IEEEPARstart{L}{arge} language models (LLM) are gaining increased attention in the machine learning community for their emerging capabilities from data and model scaling~\cite{BrownMRSKDNSSAA20,achiam2023gpt,dubey2024llama}. These models can be competitive on majorities of the text-centric tasks through post-training, e.g. instruction tuning and human preference tuning~\cite{wang2023self,ouyang2022training,rafailov2024direct}. The model after post-training shows strong behaviors to follow text instructions, which can be called instruction following capability. Recently, speech community has also spent efforts to bringing the advancement of LLM into speech-centric tasks since it is natural to connect speech and LLM to unlock the speech interactions with LLM-enabled agents. The exploration of speech-LLM can be categorized into two aspects: chat speech-LLM for real-time dialogue experience~\cite{defossez2024moshi,wang2024freeze,xie2024mini,yu2024salmonn,zhang2023speechgpt,du2023lauragpt} and understanding speech-LLM for speech understanding tasks~\cite{Ma2024AnES,salmmon24,chu2024qwen2,WuGCZZWLLRLW23,cosmic24,fathullah2024prompting,das2024speechverse}. The chat speech-LLM is built on audio tokens (such as EnCodec~\cite{DefossezCSA23} or DAC~\cite{kumar2024high}) for speech generation, and the understanding speech-LLM connects speech encoder and LLM in the latent space through neural adapters. In theory, the performance of chat speech-LLM on speech understanding tasks would not be as competitive as understanding speech-LLM because there exists information loss by using discrete tokens. In this work, we focus on the understanding speech-LLM with text output and refer to speech-LLM as the understanding speech-LLM without further clarifications.

The most common way to achieve speech-LLM is to integrate a speech encoder into LLM with a neural adapter. The neural adapter can be either linear, convolution or transformer layers to map the speech encoder outputs to text embedding space of the LLM. For example, the authors in~\cite{Ma2024AnES,Bai2024SeedASRUD} explore such architectures for a single automatic speech recognition (ASR) task and show competitive performance compared to the expert ASR model. Later on, several works, such as Qwen2-audio~\cite{chu2024qwen2}, SALMONN~\cite{salmmon24},  SpeechVerse~\cite{das2024speechverse}, and WavLLM~\cite{wavllm24}, scale the number of tasks through instruction tuning. The instruction data is created by formulating various speech and audio centric tasks into a generative manner. Such instruction-tuned speech-LLM can do well on the seen tasks in training, while the performance on the out-of-domain/unseen tasks can be terribly bad. The model is sometimes not able to follow the task instructions not seen in the training. In addition, the above-mentioned methods will tune the LLM parameters during the instruction tuning stage for better performance on different tasks, which inevitably destroys the text capability of the LLM. It is interesting to explore how to unlock the speech understanding capabilities of LLM without hurting its text performance and this work falls into the category.

As we mentioned, the LLM after post-training shows strong instruction-following capability. Freezing the LLM parameters can be an efficient way to preserve the instruction following capability of the LLM. In fact, there are several works already studying the instruction-following speech-LLM without tuning LLM. In AudioChatLLama~\cite{Fathullah2023AudioChatLlamaTG}, the LLM response of the audio transcription is used to train speech encoder and adapter, and the model is able to follow audio instructions. However, it is unknown whether the model can follow text instructions and whether it achieves good enough performance on speech understanding tasks. In BLSP~\cite{Wang2023BLSPBL}, the authors propose behavior alignment to train speech-LLM with the LLM responses to the continual writing/repeat prompts of the ASR transcriptions. The behavior alignment can provide speech-LLM zero-shot capabilities for speech and language understanding tasks, but the performance of these tasks are not as competitive as the instruction-tuned model.~\cite{Lu2024DevelopingIS} and ~\cite{kang2024frozen} further propose to add emotion and para-lingual information when generating LLM response so that the para-lingual information can be preserved in speech encoder. \cite{deng-etal-2025-wav2prompt} reveals the similar zero-shot behaviour with ASR training data only. Overall, the prior methods unlock the instruction-following speech-LLM through data efforts and do not consider a well-aligned speech and text joint space except~\cite{deng-etal-2025-wav2prompt}. Consequently, the model performance would be limited. Moreover, we are lacking a systematic way to track the instruction following capability of the speech-LLM, and thus it is hard to evaluate the effectiveness of the instruction-following speech-LLM, especially when the model fails to follow the instructions.

In this work, to evaluate the instruction-following capability of the speech-LLM, we propose to use instruction following rate (IFR) as an additional metric to evaluate the effectiveness of the instruction-following speech-LLM and offer a systematic view of calculating IFR. We study the zero-shot instruction-following speech-LLM with the Phi3.1-mini-instruct model~\cite{abdin2024phi,abouelenin2025phi}. Interestingly, we find that audio-first inputs (audio embedding + text instruction) always achieves higher IFR than instruction-first inputs (text instruction + audio embedding). This is mainly due to the reason that the Phi3 model is post-trained with instruction-first data so that audio-first training can be regarded as learning from various prompts, preventing the overfitting to the single prompt in the prefix. More importantly, we propose a novel neural adapter, AlignFormer, between the speech encoder and LLM for better speech and text modality matching. AlignFormer consists of CTC~\cite{GravesFGS06} and dynamic-window QFormer~\cite{salmmon24} layers to carefully handle the length mismatch between the speech and text sequences. Concretely, the CTC layer would provide the length and alignment information between the speech and text sequence~\cite{cassnatFan21,fan2024unienc}. The dynamic-window qformer layers would leverage the information to generate speech outputs with length and embedding space to be similar to that of the text sequence. We show for the first time the zero-shot instruction-following speech-LLM with the ASR training data only (without LLM response data). On the evaluated translation and QA tasks, AlignFormer achieves almost 100\% IFR in the audio-first setting, and improves the averaged IFR from 15\% to 46\% in the instruction-first setting, especially the game-changing improvements from zero to non-zero IFR on Gaoka and Speaker Verification QA tasks. The emerging zero-shot capability indicates that the proposed method is possible to align the speech into the LLM embedding space, which is a big step towards the perfect modality matching between the speech and text.

The contributions of this paper are:

\begin{itemize}
    \item{We propose a novel neural adapter, AlignFormer between the speech encoder and LLM to achieve zero-shot instruction-following speech-LLM with high IFR.}
    \item{We analyze the instruction-following capability of the speech-LLM and find that the audio-first training can achieve higher IFR than instruction-first training when using only the ASR data.}
    \item{We are the first to propose using IFR to evaluate the effectiveness of instruction-following speech-LLM. The usage of IFR can provide insights into the development of instruction-following speech-LLM.}
\end{itemize}

The remainder of this paper is organized as follows. Section \ref{sec:related_work} overviews the related work to this paper. Section \ref{sec:method} describes the proposed AlignFormer and methodologies we used for achieving zero-shot capability of instruction-following speech-LLM. Experimental setups are described in Section \ref{sec:exp_setup}. Results are shown and discussed in Section \ref{sec:result}. We conclude the paper in Section \ref{sec:conclusion}.

\section{Other Related Work}
\label{sec:related_work}

\textbf{CTC Compressor:} The proposed AlignFormer is closely related to CTC compressor~\cite{GaidoCNT21,WuGCZZWLLRLW23}. In fact, The AlignFormer can be regarded as a variant of CTC compressor. The difference is that we merge the blank tokens corresponding frames into non-blank tokens when reducing the speech sequence lengths and adopt the attention layers~\cite{ctcgmmrui24} to merge frames for each token instead of using mean vectors. We assume that preserving blank tokens may impede the perfect modality matching between speech and text. Another significant difference compared to~\cite{WuGCZZWLLRLW23} is that~\cite{WuGCZZWLLRLW23} can only work on ST tasks when hooking up the CTC compressor with LLM by using ST training data, while our work is to explore the zero-shot instruction following capabilities by using only ASR data.

\textbf{Speech and Text Joint Modeling:} Speech and text joint modelling is a long-standing research topic in the speech area. For example, end-to-end training of ASR and ST systems would utilize only the paired (speech, text) training data which are expensive to obtain. It would be interesting to explore the potential of leveraging data from text domain to improve the performance. Modality matching is an ideal solution to the joint modeling of speech and text. Many works have been done on joint speech-text modeling. SpeechT5~\cite{ao2022speecht5} and speechLM~\cite{zhang2024speechlm} align the speech and text modalities with mix-up or swapping strategy. In~\cite{barrault2023seamless}, length adaptor is applied to minimize the length discrepancy between speech and text, similar to M-adaptor~\cite{zhao2022Madaptor} which uses multi-head pooled self attention. Maestro ~\cite{chen22MAESTRO} learns shared representation for speech and text in RNN-T framework to benefit ASR. CJST~\cite{Zhou2024CJSTCC} uses CTC compressor for joint speech and text joint modeling.

\section{Methodologies}
\label{sec:method}

In this section, we introduce the methodologies to achieve zero-shot instruction-following speech-LLM. We first introduce the basic architecture of the speech-LLM and different audio positions for LLM inputs. The proposed AlignFormer is then introduced in details. The section ends with how we create the evaluation data for calculating the metric of IFR.

\subsection{Architecture Basics: Speech-LLM and Audio Positions}
\label{ssec:method-speechllm}

\begin{figure*}[t]
\centering
\subfloat[\footnotesize{The diagram of Speech-LLM}]{\includegraphics[width=0.6\textwidth,height=0.30\textwidth]{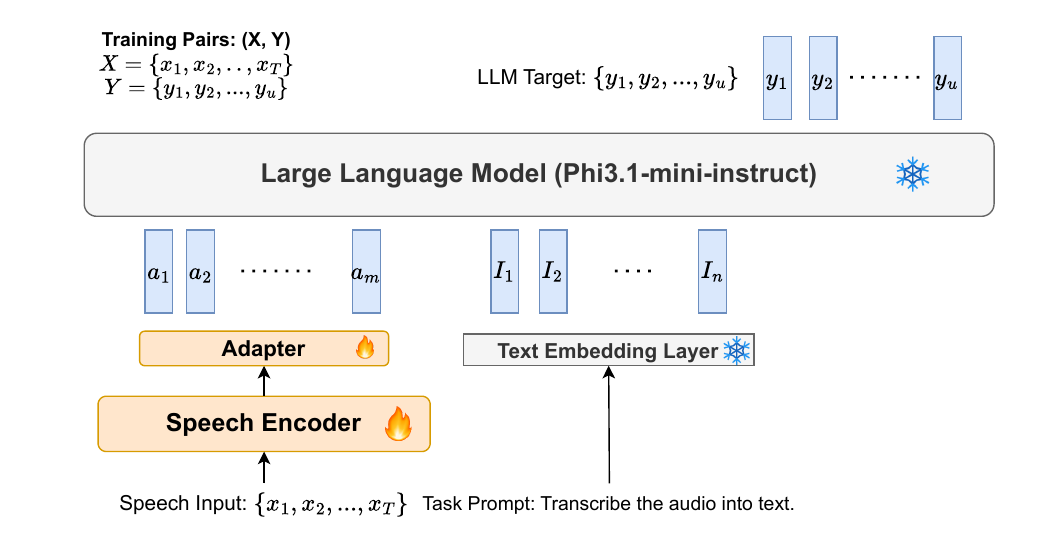}} %
\quad
\subfloat[\footnotesize{Examples of the Audio Position}]
{\includegraphics[width=0.48\textwidth,height=0.28\textwidth]{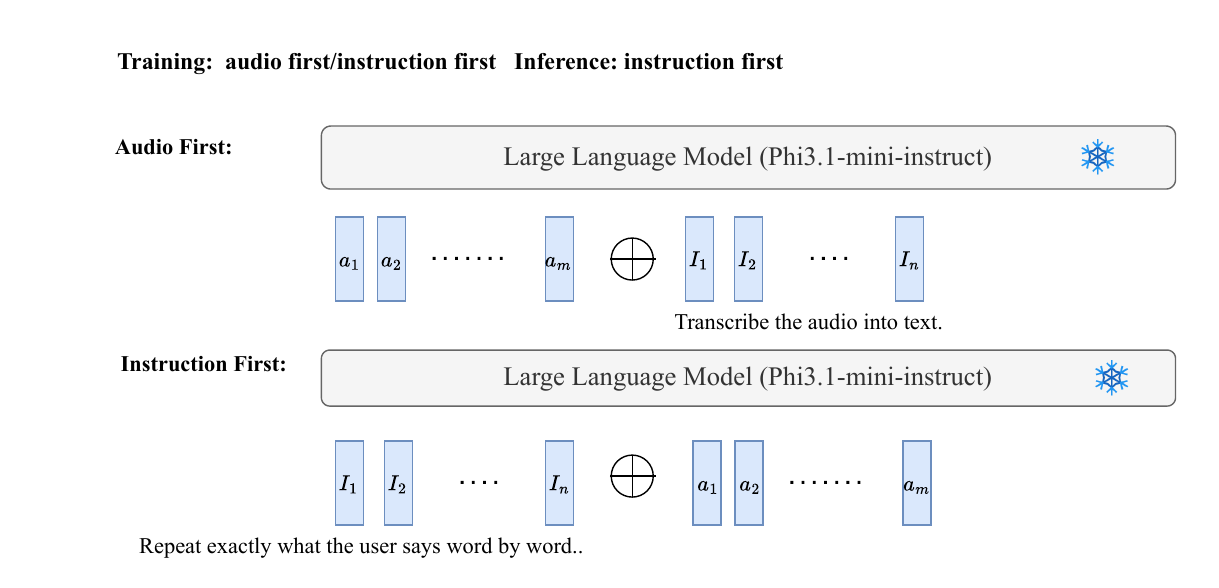}} %
\subfloat[\footnotesize{A novel Adapter: AlignFormer}]
{\includegraphics[width=0.48\textwidth,height=0.30\textwidth]{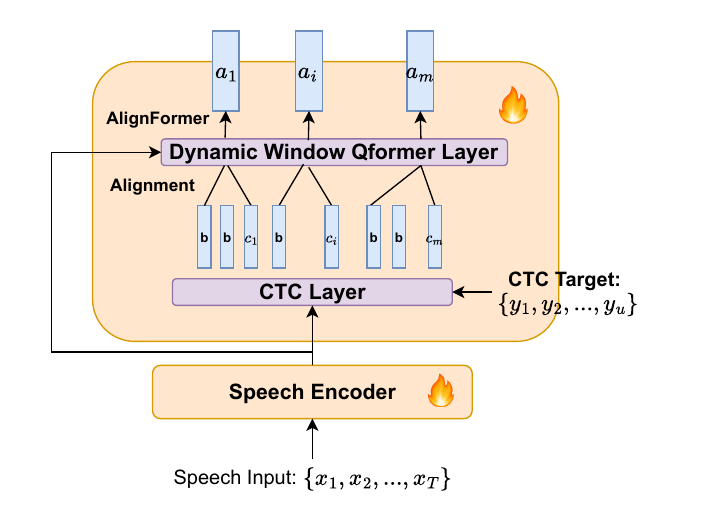}} %
\caption{(a) The architecture of speech-LLM with the ASR task prompt. (b) Different organizations of the LLM input embeddings: audio first or instruction first during training. We use instruction-first template during the inference for zero-shot evaluations because the Phi3.1-mini-instruct model is trained with instruction-first text data in the post-training stage. (c) The proposed AlignFormer module as a replacement of the adapter in speech-LLM to obtain audio embeddings $\{a_1, a_i, ..., a_m\}$.}
\label{fig:speechllm-basic}
\end{figure*}

The most common solution for building speech-LLM is to connect speech encoder and LLM with an adapter as shown in Figure~\ref{fig:speechllm-basic}a. Given a training pair $(X, Y)$, where $X$ represents the speech input sequence $(x_1,x_2,…,x_T)$ and $Y$ is the text output sequence $(y_1,y_2,…,y_U)$, the speech encoder generates high level speech representations from speech inputs $X$. Then, the adapter converts the speech representations to speech embeddings $A=\{a_1, a_2,...,a_m\}$ with the dimension of the LLM embedding space, where $m<T$ because of the sub-sampling layer in the speech encoder and/or adapter. To enable speech-LLM the instruction following capability, the task prompt $I$ is incorporated so that the speech embedding can be concatenated with the prompt embeddings as the LLM inputs. The output sequence $Y$ is used for next token prediction training for the decoder-only LLM, formulated as:

\begin{equation}
  L_{\text{ntp}} = \sum_{n=1}^{U-1}logP(y_{n+1}|y_{n}, A, I)
  \label{eq:loss_ntp}
\end{equation}
where $I$ is the task prompt. $I$ can be different for different speech and audio tasks. The speech-LLM can learn to follow these prompts through the instruction tuning stage. To achieve better speech understanding performance, LLM parameters are either full-fintuned or through LoRA finetuning~\cite{loraHuSWALWWC22}. However, the text capability, especially the instruction following capability is destroyed. The speech-LLM might only follow the text instructions seen in the training stage. The goal of this work is achieving speech and text modality matching so as to unlock the instruction following capability of LLM. We, therefore, freeze the LLM parameters in the training stage.

When organizing the speech and prompt embeddings, there are two choices: 1) audio first which can be formulated as $A\oplus I$; and 2) instruction first formulated as $I\oplus A$, as shown in Figure~\ref{fig:speechllm-basic}b. Although speech-LLM has been frequently explored in literature, there seems no paper discussing the effect of audio position and both the choices have been ever used in previous works. For example, Qwen2-audio~\cite{chu2024qwen2} and SALMONN~\cite{salmmon24} adopted the audio-first LLM inputs. BLSP~\cite{Wang2023BLSPBL} and speech-LLaMA~\cite{WuGCZZWLLRLW23} used instruction-first inputs. The reason can be that the audio position does not affect the performance of training tasks as long as they are matched during inference. However, we find that the audio position would affect the zero-shot performance for instruction-following speech-LLM and we will discuss it in Section~\ref{ssec:result_audio_pos}.

\subsection{AlignFormer: Speech and Text Modality Matching}
\label{ssec:method_alignformer}
Existing solutions for adapter is using either linear, convolution, or attention layers. Although there are down-sampling strategies in the adapter to reduce the length of the speech sequence, the length gap between the speech and text sequence still exists. Motivated by~\cite{cassnatFan23J} and~\cite{ctcgmmrui24}, we propose a novel adapter module, AlignFormer (as shown in Figure~\ref{fig:speechllm-basic}c), to handle the length mismatch between the speech and text sequences. As shown in the figure, the AlignFormer module consists of CTC and dynamic-window QFormer layers. CTC~\cite{GravesFGS06} calculates the sequence-level alignment loss between the speech inputs $X$ and ASR transcription $Y$, and is a widely used ASR loss. Each path in CTC output can provide two types of information: 1) non-blank tokens length $m$, and 2) frames coverage of speech encoder outputs for each non-blank token $c_i$. The two information can be regarded as the alignment between the non-blank token sequence and speech input sequence. Note that the length of non-blank token length ($m$) can be different from that of the target sequence length $U$. The alignment information is converted attention mask in QFormer as similar in~\cite{cassnatFan23J}, where the blank-token corresponding frames are merged to the first future non-blank token. The dynamic-window QFormer layers gather the frames of speech encoder outputs for each non-blank token to obtain the aligned embeddings $A=(a_1,a_2,…,a_m)$. The aligned embeddings are then concatenated with the instruction embeddings $I=(I_1,I_2,…,I_n)$ to form the LLM input embeddings with length of $m + n$. The entire model is trained with next token prediction loss (ASR transcription as the target) and the CTC loss. 

\begin{equation}
  L_{\text{align}} = L_{\text{ntp}} + \lambda L_{\text{CTC}}
  \label{eq:align}
\end{equation}

We explore different CTC alignments in this work and study how they affect the modality matching quality.
\begin{itemize}
    \item Greedy-alignment: The CTC greedy search path is used to get the alignment information between the speech encoder outputs and non-blank tokens, where $m$ is not guaranteed to equal $U$ because of potential insertion and deletion errors.
    \item Forced-alignment: The ASR ground-truth is used to obtain the most probable path in CTC outputs, where $m=U$. We assume forced-alignment can achieve more strict modality matching between the speech and text sequences.
    \item Mixed-alignment: Since forced-alignment is not available during inference, there would be mismatch between training and inference when using forced-alignment. We, therefore, try to mix the alignments in training. At the early stage of the training, only the forced-alignment is used. Later on, greedy-alignment is mixed in training with a probability of $p_{greedy}$.  
\end{itemize}

\subsection{IFR: Metric and Evaluation Paradigm}
\label{ssec:method_ifr}
When studying the instruction-following speech-LLM, we find that the instruction following capability is not stable, aka. some instructions are followed while others are not. It would be hard to track the effectiveness of the instruction-following speech-LLM with only the task specific metrics, such as word error rate (WER) for ASR, BLUE scores for speech translation (ST) and accuracy for speech question answering (SQA) tasks. The low scores may come from the model not following the instructions. We, therefore, propose to use IFR to track the instruction following capability of a speech-LLM model. The IFR is simple to calculate as the ratio of number of followed examples to the total examples in the evaluated dataset.

The difficulty of calculating IFR is how to detect if the answer is followed by the request in the prompt. To solve the problem, we attempt to create different prompts for each task and explicitly instruct the model answering in a specific format. Taking the emotion recognition task as an example, we provide the emotion categories in the prompt and inform the model to response with one of the options in a specific format. We detect the format in the answer to compute the IFR and accuracy for the emotion recognition task. Most of the close-ended SQA tasks can be evaluated in such a manner. For ST tasks, we detect the target language to see if the model responses in the correct language. We don't evaluate IFR for the open-ended SQA and ASR tasks because it is hard to detect if the model follow these instructions. The prompts used for different tasks are listed in Table~\ref{tab:prompts}.

\section{Experimental Settings}
\label{sec:exp_setup}
In this section, we introduce the experimental settings of this work, including the data, model, training and evaluation.

\subsection{Training Settings}
\label{ssec:train_setup}

\subsubsection{Data and Model}
\label{sssec:train_data_model}

We use the 38k hours anonymized in-house English ASR data to train the speech-LLM. For audio-first training, the task prompt is ``Transcribe the audio clip into text.". For instruction-first training, we use ``Repeat exactly what the user says word by word." as the task prompt. The reason we use this task prompt is that the LLM backbone (Phi) can follow the instruction well, where near 90\% of the LLM response to the ASR transcription is with WER lower than 5\%. We use the internal zero-shot TTS system to convert the text instruction to audio instruction when necessary.  

We use the Phi3.1-mini-instruct~\cite{abdin2024phi} model\footnote{\url{https://huggingface.co/microsoft/Phi-3-mini-128k-instruct}} as the LLM backbone in this study. Phi3.1 model is a 3.8 billion small language model, but can achieve comparable performance to 7B models like LLaMA~\cite{dubey2024llama}. The speech encoder consists 3 convolutions layers and 24 conformer blocks~\cite{conformer20} with 1024 attention dimension, 1536 feed-forward dimension, and 16 attention heads. The convolution layers contribute to a sub-sampling rate of 8, and thus 80ms token rate for adapter. Architecture-wise, we train two baseline models. The first baseline is a 2-layer MLP as the adapter without further downsample. The token rate is 80ms. The second baseline is incorporating additional window qformer layer between speech encoder and the MLP with a window size of 4. Each 4 speech encoder outputs will be further grouped into one embeeding, leading to 320ms token rate. This is a more fair comparison to AlignFormer as its token rate is about 320ms. For AlignFormer, it is 2-layer cross-attention block with one learned query. Either greedy-alignment, force-alignment, or mixed alignment is used in training. The CTC alignment is converted to attention mask functioned as the dynamic window size. The CTC task ratio is 0.3. The model parameters of speech encoder and 2-layer MLP is 450M and it is 500M for speech encoder plus AlignFormer.

\subsubsection{Training Details}
\label{sssec:training_setting}

The speech encoder is initialized from the encoder of a well-trained attention-based encoder decoder (AED) model. The LLM is initialized from Phi3.1, and thus only adapter is randomly initiated. For AlignFormer, the CTC layer is initialized from AED model as well but trained with Phi3.1 tokenizer. We show how the CTC layer initialization affects the performance in Section~\ref{ssec:result_alignformer_ablation}. All models are trained with 14000 steps with a batch size of 200k tokens, including speech and text tokens. For mixed-alignment training, we use forced-alignment in the first 7k steps and the greedy-alignment probability $p_{greedy}$ is linearly increased to 0.5 at the end of training. The training covers about 42k hours ASR data which is about the one sweep of the training data. Each training takes about 1.5 days with 8 A100 GPUs. We use the AdamW and linear decay scheduler with a peak learning rate of 4e-5 and warm up the learning rate in the first 100 steps. Deepspeed ZeRO~\cite{rajbhandari2020zero} stage 1 is used in training.

\subsection{Evaluation Settings}
\label{ssec:eval_setup}

\subsubsection{Data}
\label{sssec:eval_data_setting}

\begin{table}[tp]
\centering
\scriptsize
\caption{A summary of the metrics and instructions used for evaluation. For In-house ASR task, the prompt (instruction and audio position) is the same as what is used in training to track if the model is trained as expected. WavLLM and EN-X ST evaluations are all in a zero-shot manner. Hence, the instruction-first prompt is used. In Gaokao (listening comprehension) QA, three choices (A/B/C) are offered in the prompt.}
\begin{tabular}{l c p{1cm} p{3.6cm}}
\toprule
Task & Metric & Dataset & Prompt\\ 
\midrule\midrule
ASR & WER & In-house Eval & The same as training.\\
\midrule\midrule
\multicolumn{4}{c}{Evaluation Data from WavLLM~\cite{wavllm24}}\\
\midrule\midrule
ASR & WER & Librispeech & Repeat exactly what the user says word by word.\\

\midrule

\multirow{2}{*}{ST en-de} & BLEU & CoVoST2 & Based on the context, Translate the audio clip into German. \\ 
\cmidrule(r){2-4}
~ & BLEU & MuST-C & Based on the context, Translate the audio clip into German. \\ 
\midrule

\multirow{3}{*}{SQA} & ACC & Emotion-IEMOCAP & Based on the context, identify the emotion from the following categories: neutral, happy, sad, angry. The answer format is `The answer is: '. \\ 
\cmidrule(r){2-4}
 & ACC & Speaker Verification-Voxceleb & Based on the context, identify if there is more than one speaker. Please answer `yes' or `no'. The answer format is `The answer is: '.\\ 
\cmidrule(r){2-4}
 & ACC & Gaokao & Based on the context, answer the questions with the choice A/B/C. The answer format is `The answer is: '. \{multiple choices\}.\\
\midrule\midrule
\multicolumn{4}{c}{General ST Evaluation EN-X}\\
\midrule\midrule
\multirow{2}{*}{ST} & BLEU &CoVoST2 & Translate the audio to \{tgt\_lang\}.\\
\cmidrule(r){2-4}
 & BLEU & FLEURS & Translate the audio to \{tgt\_lang\}.\\
\bottomrule

\label{tab:prompts}
\end{tabular}
\end{table}

We use two major data sources to evaluate the performance of the instruction-following speech-LLM. One data source is from in-house evaluation. We only evaluate the ASR performance on this data to track if the model training is ongoing as expected. The task prompt is the same as used in training. The other data source is from WavLLM~\cite{wavllm24}. It contains the ASR task using the Librispeech test data, speech translation (ST) task using CoVoST2 and MuST-C en-de test data, and speech question answering (SQA) task using IEMOCAP, Voxceleb, and English Listening Comprehension examination (Chinese Gaokao) data. The SQA with Voxceleb data is tasked for speaker verification where two audios are sampled from the dataset and the model is asked whether the audios are from the same speaker. We additionally test the ST performance on CoVoST2 and FLEURS datasets in EN-X direction for more general evaluation. All the evaluations on WavLLM and ST tasks are in a zero-shot manner as only the Enlish ASR data is used in training.  

\subsubsection{Evaluation Details}
\label{sssec:eval_setting}
The metrics and instructions used for each task during evaluation is presented in Table~\ref{tab:prompts}. All prompts except in-house evaluation using an instruct-first template. This is because Phi3.1 instruct model is trained with instruction-first data. When LLM is frozen during training, we attempt to unlock the instruction following for speech inputs by speech and text modality matching. By tracking the IFR for instruction-first evaluation, it can provide us the insights into the quality of the modality matching. For ST task, we use the model\footnote{\url{https://huggingface.co/eleldar/language-detection}} in ~\cite{ConneauKGCWGGOZ20} to detect if the model answers in the target language. The greedy-alignment is used during inference. 

Note that the IFR metric should be very sensitive to prompts. However, once the prompt and detection criteria are determined, the IFR results are stable and reliable. In this work, we only evaluate IFR with the prompt shown in Table~\ref{tab:prompts}. It is interesting to evaluate with more prompt variations for general conclusions. However, creating a prompt pool that can both easily detect the following behavior is not easy. We leave it for future work.

\section{Results and Discussion}
\label{sec:result} 
\begin{table}[t]
\caption{The effect of audio position on the instruction-following speech-LLM. E1: audio-first inputs. E2: instruction-first inputs. E3: E2 + synthetic audio ASR-instruction. E4: E3 + 5 different synthetic audio ASR-instructions. $A$ is audio embeddings and $I$ is the text instruction embeddings.}
\footnotesize
\centering
\begin{tabular}{l c c c c c}
\toprule
\multicolumn{2}{c}{WavLLM Eval IFR (\%)} & E1 & E2  & E3  &  E4   \\
 & & $A\oplus I$ & $I\oplus A$  & $A\oplus A$  &  $A\oplus A$   \\

\midrule\midrule
\multirow{2}{*}{ST (en-\>de)} & CoVoST2 & 0.96 & 0.60 & 0.70 & 0.86 \\
~ & MuST-C & 0.45 & 0.07 & 0.20 & 0.80  \\
\midrule
\multirow{3}{*}{SQA} & Emotion & 1.0 & 0.07 & 0.77 & 0.56 \\
~ & Speaker Veri & 0.93 & 0.0 & 0.04 &  0.08 \\
~ & Gaokao & 0.98 & 0.0 & 0.0 & 0.0 \\
\midrule
\multicolumn{2}{c}{\textbf{Average}} & \textbf{0.85} & \textbf{0.15} & \textbf{0.34} & \textbf{0.46} \\
\bottomrule

\end{tabular}
\label{tab:audio_position}
\end{table}
In this section, we first analyze the effect of audio positions during training to the instruction-following speech-LLM. Then, we show how AlignFormer can achieve better speech and text modality matching. The ablation study of AlignFormer is also discussed.

\subsection{The Effect of Audio Positions for Unlocking Instruction-following Speech-LLM }
\label{ssec:result_audio_pos}

Before diving into the proposed AlignFormer model, we study the effect of the audio positions in training based on the speech-LLM baseline with 80ms token rate. Four experiments are conducted. 
\begin{itemize}
    \item E1: the model is trained with audio-first data.
    \item E2: the model is trained with instruction-first data.
    \item E3: the text instruction in E2 is converted to the audio version using a zero-shot TTS system.
    \item E4: E3 + 5 audio ASR instruction for diversity of the audio instructions.
\end{itemize}

The results of IFR are reported in Table~\ref{tab:audio_position}. As shown in the table, audio-first training (E1) achieves much higher IFR compared to the instruction-first training (E2). The instruction-first training can barely follow the majority of the instructions. All the failed cases as we observe are performing ASR. Since the Phi instruct model is trained with instruction-first data, we hypothesize the reason to be that the instruction-first speech-LLM training will overfit to the ASR instruction used in training. However, audio-first training is equivalent to training with large amounts of different instructions and thus the LLM instruction-following capability is preserved. 

To verify our hypotheses, we conduct the experiments of E3 and E4 to convert the text instructions into their audio version and increase the instruction diversities with 4 additional ASR prompts. The results in Table~\ref{tab:audio_position} show that by incorporating the audio ASR instruction, the averaged IFR increases from 15\% to 34\% and 46\% respectively for 1 and 5 audio instructions, showing the effectiveness of audio-first training for better instruction-following speech-LLM. However, the model still cannot follow on Gaokao QA data, indicating better speech and text modality matching alignment is required. 

\begin{tcolorbox}[colback=gray!10, colframe=black,title=Box 1. A Hallucination Example with Repeat Prompt in ASR Evaluation, label=box:audioinfo]
\textbf{Audio\_id:} 2609-169640-0017.flac

\vspace{0.5em}
\textbf{Reference:} the john behaved beautifully and came round like a top

\vspace{0.5em}
\textbf{Hypothesis:} The phrase "The John behaved beautifully. He came around like a top." is to be repeated exactly as provided, without any alterations. The text following the phrase is cut off, but the instruction is clear that the repetition should continue from the given text.

\vspace{0.5em}
\#\# Response:

The John behaved beautifully. He came around like a top.
\end{tcolorbox}

\subsection{AlignFormer: Achieving Better Speech and Text Modality Matching}
\label{ssec:result_alignformer}

\begin{table*}[t]
\caption{The zero-shot evaluations on WavLLM data. Task metrics and IFR (\%) are reported. We do not compute IFR for ASR tasks because it is hard to do so. The IFR are averaged on ST and SQA tasks. The QFormer is with 320ms token rate (window size of 4) which is close to that of AlignFormer. The token rate for E1 and E2 is 80ms. Note that the results in WavLLM paper (the second row) is with instruction tuning using in-domain instruction-following data so that it achieves 100\% IFR, while our results are with the zero-shot evaluation (only using ASR training data). CTC out + LLM indicates that the ASR greedy search results from CTC head is fed into LLM as a fair comparison to the proposed AlignFormer, which operates in the latent space. SFT-LLM-Freeze stands for freezing LLM during supervised finetuning on WavLLM speech instruction data. SFT-LLM-LoRA is experimenting with the LoRA of rank 32 on Phi3.1-mini to align with the WavLLM~\cite{wavllm24} settings.}
\footnotesize
\centering
\begin{tabular}{l  cc cc  cc cc  cc  cc c}
\toprule
\multirow{3}{*}{} & \multicolumn{2}{c}{ASR Librispeech} & \multicolumn{4}{c}{ST (en-\>de)} & \multicolumn{6}{c}{SQA WavLLM} & \multirow{2}{*}{AVG} \\
\cmidrule(r){2-3} \cmidrule(r){4-7} \cmidrule(r){8-13}
~ & test-clean & test-other & \multicolumn{2}{c}{CoVoST2} & \multicolumn{2}{c}{MuST-C} & \multicolumn{2}{c}{Emotion} & \multicolumn{2}{c}{Speaker Veri} & \multicolumn{2}{c}{Gaokao} & ~  \\
\cmidrule(r){2-3} \cmidrule(r){4-5} \cmidrule(r){6-7} \cmidrule(r){8-9} \cmidrule(r){10-11} \cmidrule(r){12-13} 

~ & WER$\downarrow$ & WER$\downarrow$ & BLEU$\uparrow$ & \textbf{IFR}$\uparrow$  &  BLEU$\uparrow$ & \textbf{IFR}$\uparrow$ & ACC$\uparrow$ & \textbf{IFR}$\uparrow$ & ACC$\uparrow$ & \textbf{IFR}$\uparrow$ & ACC$\uparrow$ & \textbf{IFR}$\uparrow$ & \textbf{IFR}$\uparrow$ \\

\midrule\midrule
Whisper + LLM~\cite{wavllm24} & 2.7 & 5.2 & 18.2 & 1.0 & 11.5 & 1.0 & - & - & - & - & 59.3 & 1.0 & - \\
WavLLM~\cite{wavllm24} & 2.22 & 5.29 & 23.99 & 1.0 & 22.14 & 1.0 & 71.96 & 1.0 & \textbf{90.21} & 1.0 & 65 & 1.0 & 1.0 \\
\midrule\midrule
E1-audio-first & 36.28 & 44.89 & 14.93 & \textbf{0.96} & 8.44 & \textbf{0.45} & 39.00 & \textbf{1.0} & 46.65 & \textbf{0.93} & 43.75 & \textbf{0.98} & \textbf{0.86} \\
\hspace{2mm} + QFormer & 4.6 & 6.19 & 11.93 & \textbf{0.74} & 1.67 & \textbf{0.13} & 9.35 & \textbf{0.32} & 0.12 & \textbf{0.02} & 1.2 & \textbf{0.02} & \textbf{0.25} \\
\hspace{2mm} + AlignFormer & 3.52 & 6.47 & 14.76 & \textbf{0.99} & 15.44 & \textbf{0.99} & 31.18 & \textbf{1.0} & 50.15 & \textbf{1.0} & 32.45 & \textbf{0.99} & \textbf{0.99} \\
\midrule 
E2-instruct-first & 3.14 & 5.91 & 11.96 & \textbf{0.60} & 1.22 & \textbf{0.07} & 2.10 & \textbf{0.07} & 0.08 & \textbf{0.0} & 0.0 & \textbf{0.0} & \textbf{0.15} \\
\hspace{2mm} + QFormer & 3.28 & 6.08 & 1.81 & \textbf{0.02} & 0.75 & \textbf{0.01} & 0.24 & \textbf{0.0} & 0.15 & \textbf{0.0} & 0.0 & \textbf{0.0} & \textbf{0.01} \\
\hspace{2mm} + Alignformer & 5.43 & 9.3 & 9.73 & \textbf{0.75} & 3.69 & \textbf{0.36} & 16.28 & \textbf{0.75} & 6.68 & \textbf{0.14} & 19.00 & \textbf{0.44} & \textbf{0.49} \\
\midrule\midrule
\multicolumn{14}{c}{E1-audio-first + AlignFormer} \\
\midrule
CTC out + LLM & 6.39 & 9.17 & 20.42 & 0.99 & 19.63 & 0.99 & 45.04 & 1.0 & 48.92 & 1.0 & 60.05 & 0.99 & 0.99 \\
+ SFT-LLM-Freeze  & 2.43 & 5.0 & 26.42 & 1.0 & 22.18 & 1.0 & 66.08 & 1.0 & 71.49 & 1.0 & 63.3 & 1.0 & 1.0 \\
+ SFT-LLM-LoRA  & 2.25 & \textbf{4.85} & \textbf{28.14} & 1.0 & \textbf{25.45} & 1.0 & 70.27 & 1.0 & 72.35 & 1.0 & \textbf{78.55} & 1.0 & 1.0 \\

\bottomrule
\end{tabular}
\label{tab:alignformer}
\end{table*}

Based on the audio-first (E1) and instruct-first training (E2), we further conduct experiments of window qformer as a fair comparison to the proposed AlignFormer since they both reduce the token rate as an adapter. The window size for qformer is 4, leading to a 320ms token rate which is close to that of AlignFormer for LLM inputs. The zero-shot results of WavLLM data are reported in Table~\ref{tab:alignformer} including the task metrics and the IFR (except ASR on Librispeech clean and other data). The results of cascaded whisper + LLM and those in WavLLM paper are presented in the table for reference. Note that the WavLLM results in the second row is with instruction tuning stage using in-domain training data. 

For the ASR tasks, we use the repeat prompt during evaluation. The poor performance of E1 on Librispeech is due to the hallucinations in addition to the ASR transcriptions. An example of such hallucinations is shown in Box 1. In other words, the model cannot follow the exact prompt to repeat and tends to speak more. This behavior is typically observed in LLM. Such hallucinations are hard to quantify so that we don't calculate IFR for the ASR tasks. Nevertheless, the ASR evaluations on audio-first training are still in a zero-shot manner and E1 should have a low IFR. For ST and SQA tasks, we can easily calculate the IFR using the method in Section~\ref{ssec:method_ifr}. We can observe from the table that AlignFormer can greatly improve the IFR for these tasks. For audio-first training, AlignFormer achieves a near 100\%. For instruction-first training, AlignFormer improves the IFR to an average of 49\%. We should specifically notice that the IFR on Gaokao QA has been improved to a non-zero IFR which is a game-changing improvement. This can be an important step towards the perfect speech and text modality matching in the LLM space and we show that it might be possible to achieve. Quality-wise, the performance of AlignFormer still encounters information loss problem because of lower token rate. For example, on CoVoST2 en-de evaluation, the BLEU scores are lower than the baselines even the IFR is higher. We still see the AlignFormer is promising because on MuST-C, the AlignFormer achieves a better BLEU than the cascades ASR + LLM system.

There is an interesting finding on the SQA tasks. For the emotion tasks, there are four categories. The accuracy of zero-shot emotion evaluation shows that the model can infer the emotion from the ASR transcriptions to some degree (better than random guess with 25\% for 4 categories). However, for the speaker verification task, the accuracy is similar to random guess (yes or no), indicating that the model is hard to infer speaker information from text. It provides us insights into the model development for emotion and speaker verification tasks.

When AlignFormer provides limited para-lingual information for LLM (e.g. random guess on the speaker verification task), a natural question is why not feeding the ASR output from CTC head to LLM for 100\% IFR. In fact, the advantage of AlignFormer is the structure flexibility of learning over latent space instead of discrete token space with ASR outputs. The para-lingual information in the audio is able to be retrieved again from the latent space through task-specific training. To verify the hypothesis, we first evaluated AlignFormer (E1-audio-first) on WavLLM test data by feeding the CTC ASR output to LLM. Later, we conducted an additional supervised finetuning (SFT) stage with the WavLLM~\cite{wavllm24} instructional training data to show the advantage of AlignFormer operating in the latent space. During SFT, we freeze the speech encoder and CTC head when most of the SFT data provides no ASR supervision for learning CTC head. We either freeze LLM and tune the AlignFormer or add LoRA module to LLM to align with the setting in the WavLLM paper. 

The results are presented in the last three columns of Table~\ref{tab:alignformer}. We observe two advantages of AlignFormer based on the results: (1) although CTC out + LLM offers good zero-shot capability for semantic heavy ST and SQA tasks, the performance on the ASR tasks with in-domain training data is much worse than the AlignFormer. The error propagation in cascades ASR + LLM solution can be reduced by learning over the latent space in AlignFormer. On the emotion and speaker verification tasks, CTC out + LLM intuitively performs bad, similarly to the AlignFormer. (2) When the AlignFormer is further trained with WavLLM SFT data, we can see that AlignFormer achieves better performance than WavLLM in most of the tasks. The gap on Emotion task is even reduced, indicating the emotion information is obtained through the adapter. However, the speaker information is still hard to be entirely retrieved from the top speech encoder layer to be on par with WavLLM structure that has dual encoders (semantic and acoustic encoders). This might be because the speech encoder is learned with ASR supervision in the alignment stage and is freezed during the SFT stage so it contains majorly semantic information.  We believe the issue can be mitigated by incorporating some forms of speaker embeddings in AlignFormer design, which is left for future work.

Overall, we believe the advantage of AlignFormer to learn modality matching in the latent space. It can reduce the error propagation existing in ASR + LLM solution for in-domain training tasks. Additionally, the AlignFormer is flexible to finetune with instructional data to retrieve para-lingual information, while ASR +LLM solution cannot. Finally, using CTC text output as the input to LLM is a cascaded system while AlignFormer is a single model that has more potential to be optimized in an end-to-end way for more complicated speech tasks.

\subsection{More Zero-shot Speech Translation Evaluations}
\label{ssec:alignformer_ablation}

\begin{table*}[t]
\caption{The zero-shot evaluation on CoVoST2 and FLERUS EN-X direction. ``GT + Phi3.1 MT" represents using the ground-truth ASR transcription of the speech inputs for Phi3.1 machine translation. ``GT + Phi3.1 MT" is a theoretical upper-bound for the instruction-following speech-LLM with ASR-only data.}
\footnotesize
\centering
\begin{tabular}{l  c ccc  c ccccccc c}
\toprule
\multirow{2}{*}{} & \multirow{2}{*}{Metric} & \multicolumn{3}{c}{CoVoST2} & \multirow{2}{*}{\textbf{AVG}} & \multicolumn{7}{c}{FLUERS} & \multirow{2}{*}{\textbf{AVG}} \\
\cmidrule(r){3-5} \cmidrule(r){7-13}
~ & ~ & DE & JA & ZH & ~ & DE & ES & FR & IT & JA & PT & ZH &  \\
\midrule\midrule
WavLLM~\cite{wavllm24} & BLEU$\uparrow$ & 24.27 & 25.00 & 30.12 & \textbf{32.31} & 22.74 & 18.61 & 25.45 & 15.41 & 21.51 & 26.2 & 26.4 & \textbf{22.33} \\
GT + Phi3.1 MT & BLEU$\uparrow$ & 26.23	& 18.81	& 21.85 & \textbf{22.30} & 29.03 & 23.95 & 38.62 & 23.8 & 18.34 & 39.59 & 23.14 & \textbf{28.07} \\
CTC out + Phi3.1 MT & BLEU$\uparrow$ & 22.35 & 17.63 & 19.67 & \textbf{19.88} & 26.85 & 21.46 & 35.45 & 20.29 & 17.54 & 34.9 & 21.97 & \textbf{25.49} \\
\midrule\midrule

\multirow{2}{*}{E1-audio-first} & BLEU$\uparrow$ & 8.59 & 5.97 & 8.67 & \textbf{7.74} & 4.44 & 4.52 & 5.65 & 3.82 & 2.64 & 5.29 & 4.11 & \textbf{4.35} \\
\cmidrule(r){3-14}
~ & IFR$\uparrow$ & 0.72 & 0.74 & 0.83 & \textbf{0.76} & 0.23 & 0.21 & 0.20 & 0.20 & 0.23 & 0.18 & 0.28 & \textbf{0.22} \\
\midrule
\multirow{2}{*}{\hspace{2mm} + AlignFormer} & BLEU$\uparrow$ & 15.7 & 8.35 & 10.97 & \textbf{11.67} & 15.72 & 14.04 & 21.97 & 12.83 & 6.36 & 24.52 & 9.77 & \textbf{15.03}  \\
\cmidrule(r){3-14}
~ & IFR$\uparrow$ & 0.98 & 0.99 & 0.98 & \textbf{0.98} & 0.99 & 0.99 & 0.99 & 0.99 & 0.98 & 0.99 & 0.97 & \textbf{0.99}  \\

\bottomrule
\end{tabular}
\label{tab:alignformer_st}
\end{table*}

We further evaluate the instruction-following speech-LLM on a more general ST setting using CoVoST2 and FLEURS dataset based on audio-first training setup. Since we used EN ASR data only, the evaluation would be conducted on EN-X direction. The results are reported in Table~\ref{tab:alignformer_st}. We also include the results of GT + Phi3.1 in the table to show Phi3's machine translation capability with the ASR ground-truth (GT). The results can serve as the upper-bound for the AlignFormer because the translation capability is from LLM. As shown in the table, we can again see the high IFR for AlignFormer across different scenarios while the instruction following capability for the baseline is not quite stable. The BLEU scores for AlignFormer show that the model is usable in some scenarios but there are still rooms for improvements compared to the upper-bound. The translation capability in LLM has not been fully evacuated yet.

\subsection{Ablation Study of AlignFormer}
\label{ssec:result_alignformer_ablation}

\begin{table*}[t]
\caption{Ablation study of the AlignFormer. ``-ctcinit" indicates using a random initialized CTC head. Please refer to Section~\ref{ssec:result_alignformer} for the details of different alignment used in AlignFormer.}
\footnotesize
\centering
\begin{tabular}{l  cc cc  cc cc  cc  cc c}
\toprule
\multirow{3}{*}{} & \multicolumn{2}{c}{ASR} & \multicolumn{4}{c}{ST (en-\>de)} & \multicolumn{6}{c}{SQA WavLLM} & \multirow{2}{*}{\textbf{AVG}} \\
\cmidrule(r){2-3} \cmidrule(r){4-7} \cmidrule(r){8-13}
~ & In-house & LS-other & \multicolumn{2}{c}{CoVoST2} & \multicolumn{2}{c}{MuST-C} & \multicolumn{2}{c}{Emotion} & \multicolumn{2}{c}{Speaker Veri} & \multicolumn{2}{c}{Gaokao} & ~  \\
\cmidrule(r){2-3} \cmidrule(r){4-5} \cmidrule(r){6-7} \cmidrule(r){8-9} \cmidrule(r){10-11} \cmidrule(r){12-13} 

~ & WER$\downarrow$ & WER$\downarrow$ & BLEU$\uparrow$ & \textbf{IFR}$\uparrow$  &  BLEU$\uparrow$ & \textbf{IFR}$\uparrow$ & ACC$\uparrow$ & \textbf{IFR}$\uparrow$ & ACC$\uparrow$ & \textbf{IFR}$\uparrow$ & ACC$\uparrow$ & \textbf{IFR}$\uparrow$ & \textbf{IFR}$\uparrow$ \\

\midrule\midrule
E1-audio-first & 10.72 & 44.89 & 14.93 & \textbf{0.96} & 8.44 & \textbf{0.45} & \textbf{39.00} & 1.0 & 46.65 & \textbf{0.93} & 43.75 & \textbf{0.98} & \textbf{0.86} \\
\hspace{2mm} + QFormer & 12.16 & 6.19 & 11.93 & \textbf{0.74} & 1.67 & \textbf{0.13} & 9.35 & \textbf{0.32} & 0.12 & \textbf{0.02} & 1.2 & \textbf{0.02} & \textbf{0.25} \\
\midrule\midrule
\multicolumn{14}{c}{AlignFormer} \\
\midrule\midrule

greedy-alignment & 13.19 & 45.22 & 14.82 & \textbf{0.99} & 14.88 & \textbf{0.99} & 31.59 & \textbf{1.0} & 49.43 & \textbf{1.0} & 33.65 & \textbf{0.99} & \textbf{0.99} \\
forced-alignment & 11.47 & 6.6 & 13.51 & \textbf{0.99} & 13.55 & \textbf{0.91} & 29.65 & \textbf{0.97} & 48.46 & \textbf{0.99} & 35.45 & \textbf{0.98} & \textbf{0.97} \\
\hspace{2mm} - ctcinit  & 12.07 & 75.18 & 13.26 & \textbf{0.96} & 8.91 & \textbf{0.71} & 19.1 & \textbf{0.91} & 41.43 & \textbf{0.89} & 33.3 & \textbf{0.89} & \textbf{0.87} \\

mixed-alignment & 12.87 & 6.47 & 14.76 & \textbf{0.99} & 15.44 & \textbf{0.99} & 31.18 & \textbf{1.0} & 50.15 & \textbf{1.0} & 32.45 & \textbf{0.99} & \textbf{0.99} \\

\bottomrule
\end{tabular}
\label{tab:alignformer_ablation}
\end{table*}

The AlignFormer results reported in previous sections are based on the overall best setting. In this section, we show the ablation study of the AlignFormer including the choice of alignment used in training and the effect of initializing the CTC head with a well-trained layer. The results are summarized in Table~\ref{tab:alignformer_ablation}. The In-house ASR evaluation is using audio-first template to be consistent as training so the WER performance can reflect how much information will lose by more aggressive downsampling. We can see from the table that with the forced-alignment in training, the WER performance of AlignFormer (\~320ms token rate) is better than QFormer and has only 5\% relative WER degradation compared to the 80ms token rate. In other words, the proposed AlignFormer can also be used as an effective model for long-form ASR and summarization tasks by reducing the length of speech inputs.

Regarding the zero-shot evaluations on WavLLM, we see that both the greedy-alignment and forced-alignment can provide high IFR. However, their instruction following capability is not stable. For example, the model with greedy-alignment training cannot always follow the repeat prompt on Librispeech test-other data. The model with forced-alignment training does not follow some translation prompts in MuST-C data. This might be because the token length of the greedy-alignment does not necessarily equal that of the ground-truth tokens (insertion or deletion errors), resulting in imperfect modality matching. On the other hand, forced-alignment can achive perfect modality matching with the same token length in training. However, forced-alignment is not available in inference, leading to a training inference mismatch. To mitigate the issue, we experiment the mixed-alignment where forced-alignment is used in the early stage of the training and then greedy-alignment is mixed into the training. The results show that the mixed-alignment can achieve the overall best performance on all tasks, except that the in-house ASR performance is degraded.

We also show in the table the importance of CTC-head initialization when using limited data for speech and text modality matching. This is because the CTC alignment quality would affect the convergence of modality matching. When scaling the AlignFormer with more training data, the CTC-head initialization might not be that important.

Finally, we calculate the similarity between the text and speech embeddings on E1-audio-first + AlignFormer using the librispeech test-other data. The average cosine similarity on the all utterances is 0.12. which is not high. We also calculate the average cosine similarity when AlignFormer is not used, the number of which is 0.08. The analysis shows the potential to incorporate additional distance loss between the speech and text embeddings for better modality matching, which is left for future work.

\section{Conclusions}
\label{sec:conclusion}

In this paper, we studied zero-shot instruction-following speech-LLM and proposed to use instruction following rate (IFR) to evaluate the performance of instruction-following speech-LLM. To calculate IFR, we offered a systematic perspective by providing choices and the answer format in the prompt for different tasks. The IFR can be calculated by detecting the pattern in the LLM response. In addition, we analyzed the effect of audio position in training to the instruction following capability. We found that audio-first training achieved higher IFR than the instruction-first training due to a higher exposure to more diversified instructions. More importantly, we proposed a novel adapter, AlignFormer, to connect speech encoder and LLM for better speech and text modality matching. By using only ASR training data, AlignFormer can perform speech translation (ST) and speech question answering (SQA) tasks in a zero-shot manner. On the zero-shot evaluation of these tasks, AlignFormer achieved nearly 100\% IFR in the audio-first setting and improved IFR from 15\% to 49\% in the instruction-frist setting. On a more general ST evaluation, we achieved average BLEU scores of 11.67 and 15.03 on CoVoST2 and FLEURS en-X translation, respectively. Our results showed that it is possible to align the speech into LLM embedding space and presented the advantage of AlignFormer over CTC ASR + LLM cascades solution. We believe that this study is a big step towards the perfect speech and text modality matching.

\bibliographystyle{IEEEtran}
\bibliography{ref}

\vfill

\end{document}